\def\ump{\noindent}
\def\Q{Q^2}
\def\xq{(x,Q^2)}
\def\MiN{\negTwo-\negTwo} 
\def\negOne{       \mskip -1.0mu}
\def\negTwo{       \mskip -2.0mu}
\def\negFour{      \mskip -4.0mu}
\def\Half{{1\negTwo/2}}
\def\mFour{   \noalign{\vskip .4cm}}
\def\stp{^2\negFour+}
\def\b{\over\log\negTwo\left(\Q\negOne/\negOne0.04\right)}
\documentstyle[prl,aps,epsfig]{revtex}

\begin{document}
\renewcommand{\thefootnote}{\fnsymbol{footnote}}
\vskip -1.in
\begin{flushright}
{\small
  SLAC-PUB-7927\\
  August 1998 \\
}
\end{flushright}
\begin{center}{\bf\large
 Measurements of $R=\sigma_L/\sigma_T$ for $0.03<x<0.1$ and Fit
to World Data\footnote{Work supported by
Department of Energy contract  DE--AC03--76SF00515.}}
\vskip .1in
{The E143 Collaboration \break
K.~Abe,$^{15}$
T.~Akagi,$^{11,15}$
P.~L.~Anthony,$^{11}$
R.~Antonov,$^{10}$
R.~G.~Arnold,$^{1}$
T.~Averett,$^{16,\ddag\ddag}$
H.~R.~Band,$^{18}$
J.~M.~Bauer,$^{6,\S\S}$
H.~Borel,$^{4}$
P.~E.~Bosted,$^{1}$
V.~Breton,$^{3}$
J.~Button-Shafer,$^{6}$
J.~P.~Chen,$^{16,\heartsuit}$
T.~E.~Chupp,$^{7}$
J.~Clendenin,$^{11}$
C.~Comptour,$^{3}$
K.~P.~Coulter,$^{7}$
G.~Court,$^{11,*}$
D.~Crabb,$^{16}$
M.~Daoudi,$^{11}$
D.~Day,$^{16}$
F.~S.~Dietrich,$^{5}$
J.~Dunne,$^{1,\heartsuit}$
H.~Dutz,$^{11,**}$
R.~Erbacher,$^{11,12}$
J.~Fellbaum,$^{1}$
A.~Feltham,$^{2}$
H.~Fonvieille,$^{3}$
E.~Frlez,$^{16}$
D.~Garvey,$^{8}$
R.~Gearhart,$^{11}$
J.~Gomez,$^{14}$
P.~Grenier,$^{4}$
K.~A.~Griffioen,$^{10,17}$
S.~Hoibraten,$^{16,\S}$
E.~W.~Hughes,$^{11,\ddag\ddag}$
C.~Hyde-Wright,$^{9}$
J.~R.~Johnson,$^{18}$
D.~Kawall,$^{12,\diamondsuit}$
A.~Klein,$^{9}$
S.~E.~Kuhn,$^{9}$
M.~Kuriki,$^{15}$
R.~Lindgren,$^{16}$
T.~J.~Liu,$^{16}$
R.~M.~Lombard-Nelsen,$^{4}$
J.~Marroncle,$^{4}$
T.~Maruyama,$^{11}$
X.~K.~Maruyama,$^{8}$
J.~McCarthy,$^{16}$
W.~Meyer,$^{11,**}$
Z.-E.~Meziani,$^{12,13}$
R.~Minehart,$^{16}$
J.~Mitchell,$^{14}$
J.~Morgenstern,$^{4}$
G.~G.~Petratos,$^{11,\ddag}$
R.~Pitthan,$^{11}$
D.~Pocanic,$^{16}$
C.~Prescott,$^{11}$
R.~Prepost,$^{18}$
P.~Raines,$^{10,\infty}$
B.~Raue,$^{9,\dag}$
D.~Reyna,$^{1,\partial}$
A.~Rijllart,$^{11,\dag\dag}$
Y.~Roblin,$^{3}$
L.~S.~Rochester,$^{11}$
S.~E.~Rock,$^{1}$
O.~A.~Rondon,$^{16}$
I.~Sick,$^{2}$
L.~C.~Smith,$^{16}$
T.~B.~Smith,$^{7}$
M.~Spengos,$^{1,10}$
F.~Staley,$^{4}$
P.~Steiner,$^{2}$
S.~St.Lorant,$^{11}$
L.~M.~Stuart,$^{11}$
F.~Suekane,$^{15}$
Z.~M.~Szalata,$^{1}$
H.~Tang,$^{11}$
Y.~Terrien,$^{4}$
T.~Usher,$^{11}$
D.~Walz,$^{11}$
F.~Wesselmann,$^{9}$
J.~L.~White,$^{1,\infty}$
K.~Witte,$^{11}$
C.~C.~Young,$^{11}$
B.~Youngman,$^{11}$
H.~Yuta,$^{15}$
G.~Zapalac,$^{18}$
B.~Zihlmann,$^{2}$
D.~Zimmermann$^{16}$}

{\it
\baselineskip 14 pt
\vskip 0.1cm
\vskip 0.1 cm
{$^{1}$The American University, Washington, D.C. 20016}  \break
{$^{2}$Institut f\" ur Physik der Universit\" at Basel,
  CH--4056 Basel, Switzerland} \break
{$^{3}$LPC IN2P3/CNRS,
  University Blaise Pascal, F--63170 Aubiere Cedex, France}  \break
{$^{4}$DAPNIA-Service de Physique Nucleaire,
  Centre d'Etudes de Saclay, F--91191 Gif/Yvette, France} \break
{$^{5}$Lawrence Livermore National Laboratory, Livermore, California 94550}
\break
{$^{6}$University of Massachusetts,  Amherst, Massachusetts 01003}  \break
{$^{7}$University of Michigan, Ann Arbor, Michigan 48109} \break
{$^{8}$Naval Postgraduate School, Monterey, California 93943} \break
{$^{9}$Old Dominion University,  Norfolk, Virginia 23529} \break
{$^{10}$University of Pennsylvania,  Philadelphia, Pennsylvania
  19104} \break
{$^{11}$Stanford Linear Accelerator Center,
  Stanford, California 94309} \break
{$^{12}$Stanford University, Stanford, California 94305} \break
{$^{13}$Temple University, Philadelphia, Pennsylvania 19122}  \break
{$^{14}$Thomas Jefferson National Accelerator Facility, Newport News, Virginia 23606} \break
{$^{15}$Tohoku University, Sendai 980, Japan} \break
{$^{16}$University of Virginia, Charlottesville, Virginia 22901} \break
{$^{17}$The College of William and Mary, Williamsburg, Virginia 23187} \break
{$^{18}$University of Wisconsin, Madison, Wisconsin 53706}  \break 
\break \break
{\it To be submitted to Physics Letters} \break} 

\end{center}

\begin{abstract}
Measurements were made at SLAC of the cross section for scattering 
29 GeV electrons from carbon at a laboratory angle of $4.5^\circ$, 
corresponding to $0.03<x<0.1$ and $1.3<Q^2<2.7$ GeV$^2$. Values of 
 $R=\sigma_L/\sigma_T$  
were extracted in this kinematic range by 
comparing these data to cross sections measured at a higher beam energy by the 
NMC collaboration. The results are in reasonable agreement with pQCD
calculations and with extrapolations of the $R1990$ parameterization of 
previous data. 
A new fit is made including these data and other recent results.
\end{abstract}


\section*{Introduction}
The spin-averaged cross section for lepton nucleon scattering can be
written in terms of the two components for virtual photon absorption: the 
longitudinal cross section $\sigma_L(x,Q^2)$, and the transverse
cross section  $\sigma_T(x,Q^2).$  Alternatively this can be expressed
in terms of the structure functions
$F_2$ and $R=\sigma_L/\sigma_T$
\begin{eqnarray}
{d\sigma(x,Q^2,\epsilon)\over dE^{\prime}d\Omega} &=
 \Gamma(x,Q^2,\epsilon) [\sigma_T(x,Q^2)+\epsilon \sigma_L(x,Q^2)] \,\cr
 &=\sigma_{Mott}{2MxF_2(x,Q^2)\over \epsilon Q^2}
\left[{1 +\epsilon R(x,Q^2) \over 1+R(x,Q^2)}\right] \ump
\end{eqnarray}
where  $Q^2=4EE^\prime\sin^2(\theta/2)$ is the 
four-momentum squared of the virtual photon, $x=Q^2/[2M(E-E^\prime)]$ 
is the light-cone
momentum fraction of the struck parton, $\Gamma$ is the virtual photon flux, 
$\epsilon^{-1}=1+2(1+Q^2/4M^2x^2)\tan^2(\theta/2)$,  and
$\theta$ and $E^\prime$ are the scattered lepton scattering angle and
momentum in the lab.
$R$ is determined by making cross section measurements at
fixed $(x,Q^2)$ as a function of  $\epsilon$ by varying the beam energy and
scattering angle.
In the quark-parton model, $R$
is sensitive to the spin of the struck partons: at large $Q^2$, 
$R$ is zero for spin 1/2 quarks while at finite $Q^2$ quark transverse
momentum causes non-zero values.  In pQCD calculations of $R$ \cite{AM} the 
leading term is proportional to $\alpha_s$ times integrals over the quark and
gluon distributions, and is thus sensitive to the gluon content of the 
nucleon. At low $Q^2$ and high $x,$ target mass corrections \cite{tm} also
contribute to $R.$  Knowledge of $R$
is important for  extracting 
the unpolarized structure function $F_2$ from cross section
measurements and
the spin structure functions $g_1$ and $g_2$
from asymmetry measurements of polarized leptons on polarized nucleons.

 Previously a good parameterization was made of the world data on $R$ 
(known as $R1990$) \cite{r1990}, but the fit is limited in validity
to $x>0.07$, where input data existed. To extend our knowledge of $R$ into
the lower $x$ region (where it is needed for the extraction of $g_1$)
we made cross section measurements
in the range $0.03<x<0.1.$ This was part of SLAC experiment E143 \cite{E143}, 
whose primarily
goal was the measurement of $g_1$ for the proton and deuteron. 
 
\section*{E143 Cross Sections}
We measured the cross section for scattering of 29.1 GeV electrons 
from a 1.7 gm/cm$^2$ carbon target  at the End Station A facilities
of the Stanford Linear Accelerator Center.
Scattered electrons in the momentum range
of  6 to 25 GeV/c ($0.03\leq x \leq 0.4$) were detected in a magnetic 
spectrometer at $\theta=4.5^\circ.$ We used two threshold 
\v{C}erenkov counters and a segmented lead glass 
shower counter to identify electrons. Electron momenta were determined 
from tracking in a multi-plane hodoscope system and independently 
from the energy deposited in the shower counter.
 The system was designed to
measure the spin structure functions of the nucleons and thus could
operate at high  rates.  Details 
are given in \cite{E143}. The acceptance of the spectrometer was
calculated with a model that used  magnetic measurements and 
survey information.
Part of the acceptance of the spectrometer
was also calibrated by re-measuring the already   known cross section 
in the kinematic region $0.1<x<0.3 $.  These  cross sections are
accurately known at all values of $\epsilon$ from
a fit to $F_2^d$ by the NMC collaboration \cite{f2nmc}, a fit to the
$A$-dependence of lepton-nucleon scattering \cite{emcfit},
and previously measured values of $R$\cite{r1990}. 
The central momentum of the spectrometer was then
lowered in several steps to put scattered electrons corresponding to 
$0.03<x<0.1$  into the calibrated acceptance region. 
Small adjustments were made to the spectrometer acceptance model until the 
overlaps between
spectra with different central momentum settings were in good agreement.
The final corrections to the acceptance compared to the original
model were in the few percent range. Thus the overall normalization of
our results are tied to those of the NMC fit.
 
Absolute cross sections in the low $x$ region  for the carbon target
were obtained taking into account the residual background contaminations,
the experimental efficiency for detecting electrons, the trigger dead time, 
and the application of  radiative corrections \cite{tsai,e140}. 
The results for carbon cross sections per nucleon are shown in 
Fig. \ref{fg:sig}a~ and Table \ref{tb:sigma}. The systematic errors 
include an overall normalization uncertainty of about 2.5\% 
due the combined uncertainties in target
thickness, beam charge and  spectrometer acceptance. Other
systematic errors increased with decreasing $x,$ including
detection efficiency, the spectrometer acceptance, and
radiative corrections 
(about 3\% at low $x$, decreasing to about 1\% at the highest $x$). 
The $Q^2$ for the points in Fig. \ref{fg:sig}a~  vary approximately linearly 
with $x$ as in  Table  \ref{tb:sigma}.
The curve on Fig. \ref{fg:sig}a~ is the predicted cross section using the 
NMC fit to $F_2^d$\cite{f2nmc}, nuclear corrections\cite{emcfit}, and  
$R1990$\cite{r1990}. 
The good agreement between data and model for $x<0.1$ is an 
indication that the extrapolation of
$R1990$ to $x=0.03$ works reasonably well.

\section*{RESULTS FOR $R$}
To determine values for $R$, we used  Eq. 1 with  the E143 cross sections 
and those
of NMC \cite{f2nmc} at the same $(x,Q^2)$
values, but at much higher beam energies (higher values of 
$\epsilon$). Since our cross sections are 
normalized to NMC
cross sections at higher values of $x$, the normalization uncertainties
between the two experiments are negligible thus reducing the systematic
uncertainty on $R$.

\begin{figure}[b!] 
\vspace*{1.5in}
\hspace*{.45in}
\includegraphics{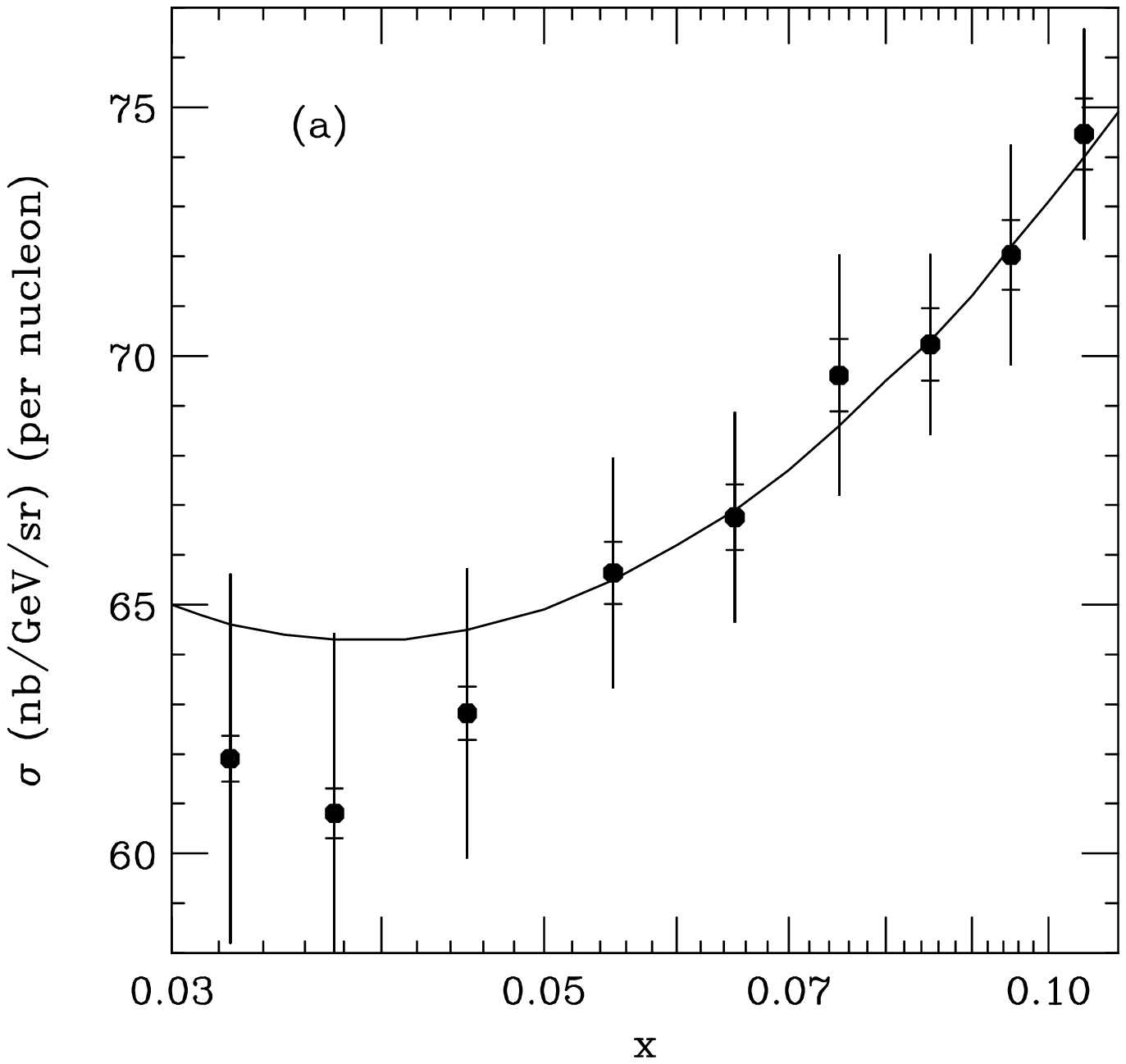} 
\includegraphics{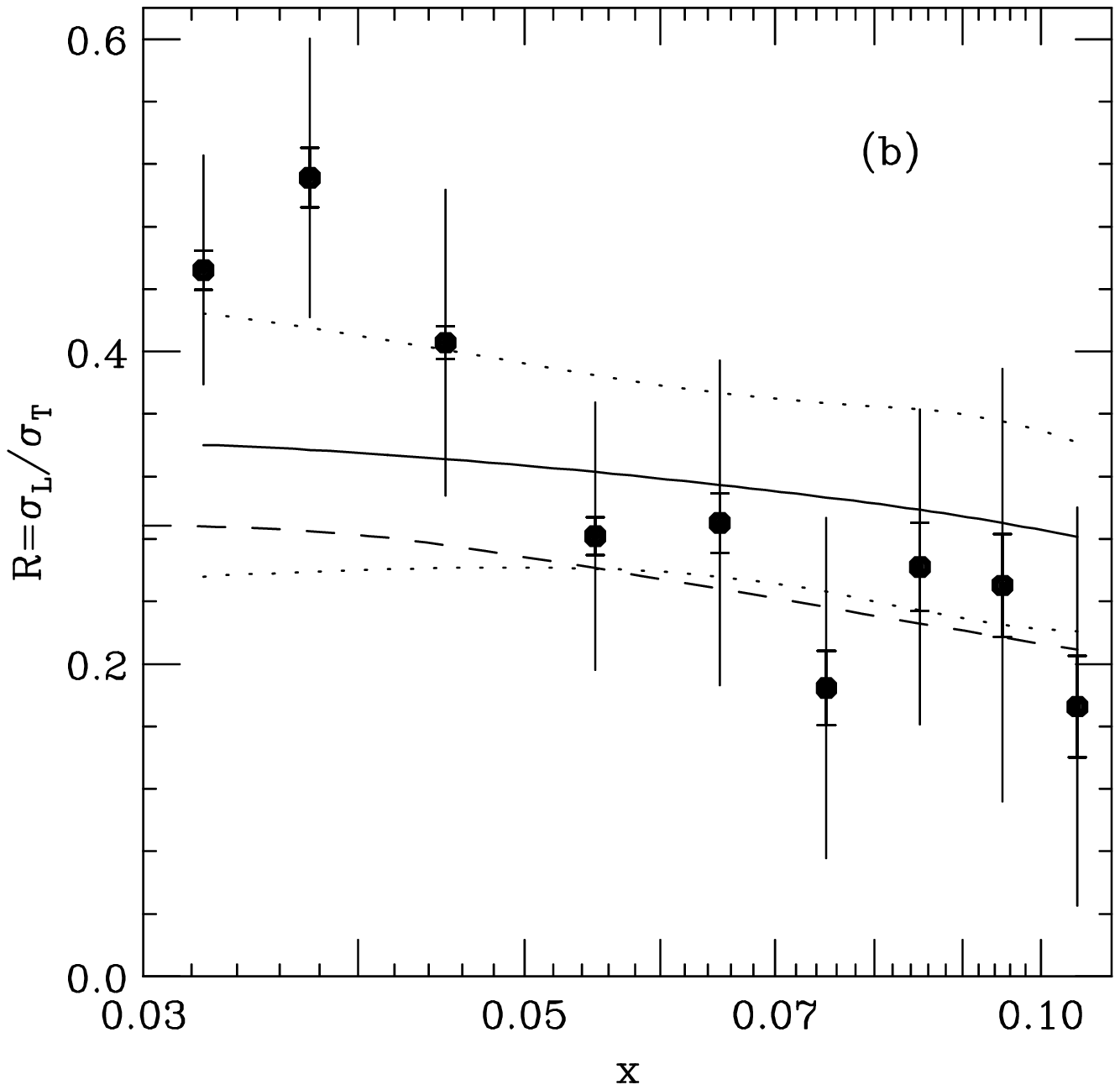}  
\vspace{110pt}   
\caption{(a) Cross sections from this experiment (E143) for 29.1 GeV electron
scattering from carbon at $4.5^\circ$. Inner (outer) 
error bars are statistical (systematic). The curve is calculated using the
NMC fit to $F_2$\protect\cite{f2nmc}, 
the R1990 fit to $R$\protect\cite{r1990}, and the $A$-dependence 
of lepton-nucleon scattering \protect\cite{emcfit}.
(b) $R$ extracted from this experiment (E143) combined with NMC.
Inner (outer) error bars are statistical (systematic).
The solid curve is the R1990 fit, with the dotted curves
showing the error band evaluated  at the $Q^2$ values of the data. The dashed
curve is a pQCD calculation described in the text. }
\label{fg:sig}
\end{figure}

The results are shown in Fig. \ref{fg:sig}b~ and Table \ref{tb:sigma} together
with various fits and calculations.
The new data are in reasonable agreement 
with  the $R1990$ parameterization (solid curve with dotted curves showing
error band), plotted at the $Q^2$ 
values of E143, although there is a
tendency for the data to be slightly higher than $R1990$
at low $x$ and lower than  $R1990$ at high $x.$ 
The lower $x$ data are also higher than a calculation of NNLO  pQCD 
plus target mass 
corrections\cite {bodek} (dashed) using the MRS-R2 \cite{mrs} parton 
distribution with four flavors.

 Since the  $R1990$ was published, there has been a
considerable body of new data on $R$ from this experiment, from  
SLAC E140X \cite{E140X}, NMC \cite{r_nmc}, and
CCFR \cite{CCFR} as well as final results
from CDHSW\cite{CDHSW}.  The data were recorded on a variety of targets,
but since $R^d=R^p=R^A$\cite{E140X,rdiff} to high accuracy we will combine 
them into a single data set.
The new data have extended the kinematic
range to lower and higher values of $x.$  Because $R1990$ has been used 
extensively outside its range of validity at low $x$, it is important to
compare it to the new data. 
The confidence level for $R1990$ to match the data
is 61\%. For the region  $x\leq 0.07$, outside the range of validity of 
R1990, the confidence level is 16\%. 
 We  note that $R_c$, one of the three
fits that were averaged to make $R1990,$ has a confidence level of less than
1\% to agree with the low $x$ data, while the $R_a$ and $R_b$ are much
more successful.

\begin{figure}
\vspace*{2.3in}  
\hspace*{.45in}
\includegraphics{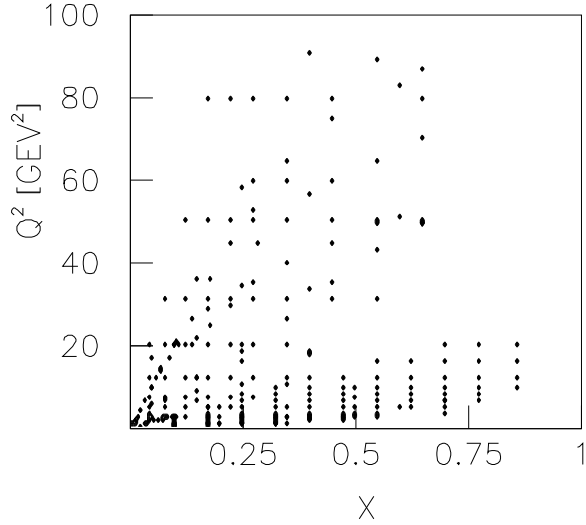}\vskip 0.7cm 
\vskip .7in
\caption{Kinematic distribution of data used for the R1998 fit.}
\label{fg:xq}
\end{figure}

We have performed a new fit
using  the  present data  \cite{r1990,E140X,r_nmc,CCFR,CDHSW}, 
but excluding data with errors greater than 0.5 
or more than 3 standard deviations below zero.
The final data set still included some values of 
$R$ which, due to errors,  were negative (unphysical).
There were 237 points with a kinematic range of $0.005\leq x \leq 0.86$
and $0.5 \leq Q^2 \leq 130.$  Figure \ref{fg:xq}~ shows the distribution
of points as a scatter plot. 
More parameters were added to the general form of the $R1990$ fits to try to
accommodate the new data at low $x.$ Three 6-parameter models were used based
on the previous three $R1990$ models:

\begin{eqnarray}
R_a&=&{a_1\b}\,\Theta\xq+{a_2\over\root4\of{Q^8+a_3^4}}[1 + a_4x +a_5x^2]
  x^{a_6}\ ,\cr\mFour
R_b&=&{b_1\b}\,\Theta\xq+[{b_2\over\Q}+{b_3\over Q^4+0.3^2}][1 + b_4x +b_5x^2]
  x^{b_6}\ ,\cr\mFour
R_c&=&{c_1\b}\,\Theta\xq+c_2\negTwo\left[
   \left(\Q\MiN\Q_{thr}\right)\stp c_3^2\,\negTwo\right]^{-\Half}
\end{eqnarray}
with
\begin{eqnarray}
\Q_{thr}  &=& c_4x +c_5x^2 +c_6x^3 \cr\mFour
\Theta\xq &=& 1 +12\big({Q^2 \over Q^2 +1}\big)\big({0.125^2 \over 0.125^2+x^2}\big),
\end{eqnarray}
where the units of $Q^2$ are GeV$^2.$
The coefficients of the fits are shown in Table \ref{tb:coef}. 
  As in the case of $R1990$, we define $R1998$ to be the
average of the three fits.
The error associated with the fitting is approximately given by:
\[ \delta R(x,Q^2)= 0.0078 -0.013x + { 0.070 -0.39x + 0.70x^2 \over 1.7 +Q^2 }.\]
This error is largest at low $Q^2,$ reaching a maximum value for $x\sim 0.04.$
A systematic error associated with the functional form can be assigned from
the spread of the three fits. Long range correlated errors such as those
due to radiative corrections will enhance the errors.
The new fits result in a  better agreement with the data, with a
confidence level of 73\% for all the data and 38\% when restricted to
$x\leq 0.07.$ 

\begin{figure}[t]
\centerline{\epsfig{file=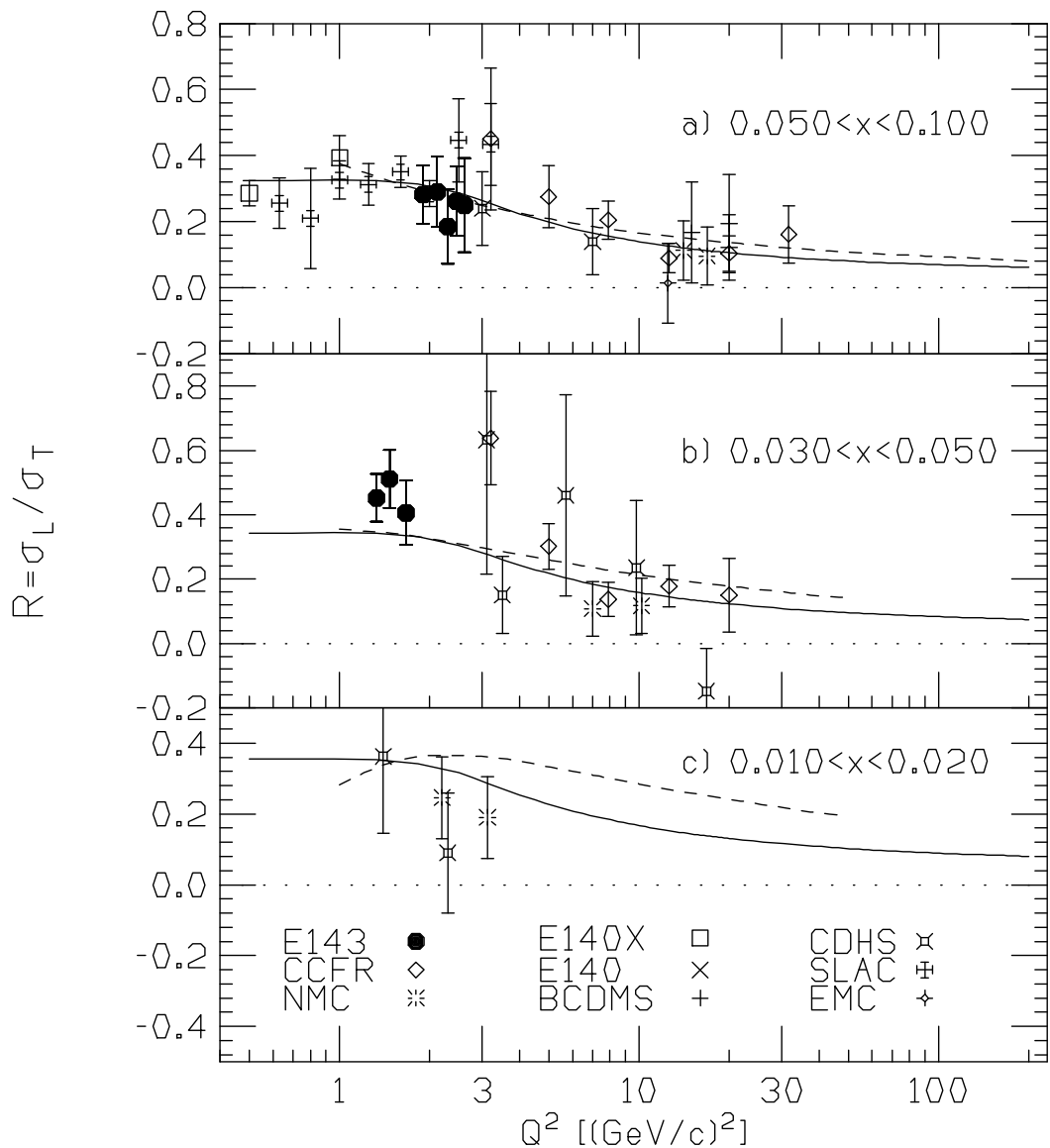,width=5.50 in}}\vskip 0.3cm
\caption{$R$ as a function of $Q^2$ for: a) $0.05\leq x \leq 0.10;$ 
 b) $0.03\leq x \leq 0.05;$
 c) $0.01 \leq x \leq 0.02.$ The solid curve is the average of the new
 fits, $R1998$, and the
dashed curve is the NNLO pQCD calculation described in the text.}
\label{fg:rx}
\end{figure}

\begin{figure}[t]
\centerline{\epsfig{file=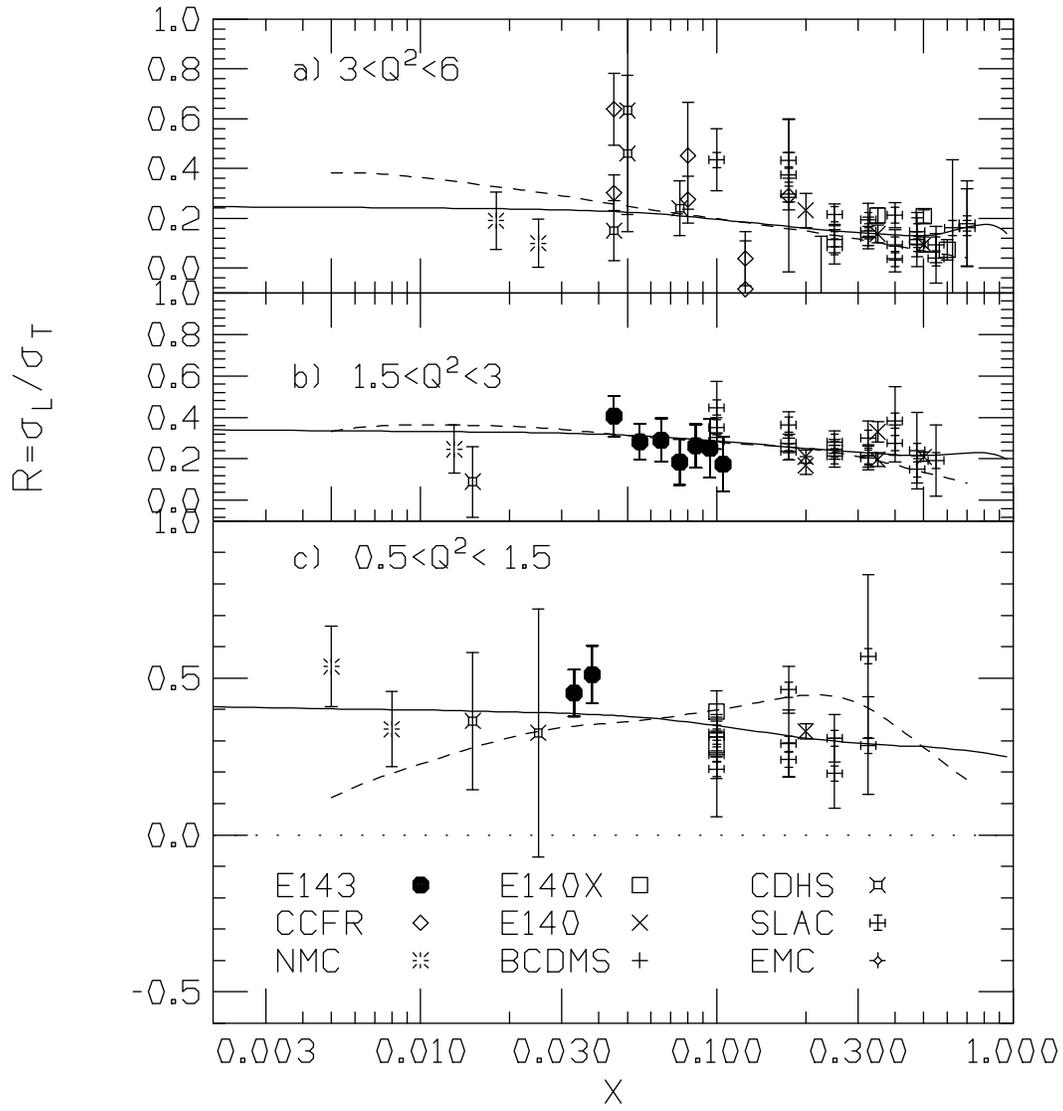,width=5.50 in}}\vskip 0.5cm
\caption{$R$ as a function of $x$ for: a) $3\leq Q^2 \leq 6$(GeV/c)$^2$;
 b) $1.5\leq Q^2 \leq 3$ (GeV/c)$^2$; 
 c) $0.5\leq Q^2 \leq 1.5$ (GeV/c)$^2$.  
The solid curve is the average of the new fits, 
$R1998,$ and the dashed curve is the NNLO pQCD calculation 
described in the text.}
\label{fg:rq}
\end{figure}

 Fig. \ref{fg:rx}~ shows the measured values of $R$ as a function of
$Q^2$ in three  ranges of $x$ below $x=0.10$, along with $R1998$ and the
pQCD plus target mass
calculation used above. The value of $R$ decreases with $Q^2$ as had been
observed \cite{e140} at higher values of $x.$  At these low values of $x,$
target mass effects make only a small contribution to the pQCD calculation.
The pQCD curve is below the data at low $Q^2$ as previously noted.
Fig. \ref{fg:rq}~ shows the data as a function of $x$ for three ranges of
$Q^2.$  $R$ is not very strongly dependent on $x$ in this $Q^2$ range, 
continuing the trend observed for $x\geq 0.07.$  The pQCD plus target mass 
calculation falls
below the data at low $Q^2$ and low $x,$ but otherwise is in quite
good agreement. We note that $Q^2 \sim 1 {\rm GeV}^2$ is a rather low value for
pQCD calculations.

In summary, our new  measurements of $R,$ as well as other recent results
are consistent with extrapolations of the empirical parameterization $R1990$. 
The result are roughly consistent with pQCD plus target mass calculations.
Our new fit to the data ($R1998$), although  similar to $R1990,$
better reflects the wealth of new data obtained over the last several years
and is in better agreement with the low $x$ data.

 We wish to acknowledge the tremendous effort made by the SLAC staff in
making this experiment successful.  This work
was supported by Department of Energy contracts: No.  W-2705-Eng-48
(LLNL), No.  DE-AC03-76SF00515 (SLAC), No.  DE-FG03-88ER40439 (Stanford),
Nos.  DE-FG05-88ER40390 and DEFG05-86ER40261 (Virginia), and No.
DE-AC02-76ER00881 (Wisconsin); by National Science Foundation Grants No.
9114958 (American), No. 9307710 (Massachusetts), No. 9217979 (Michigan),
and No. 9104975 (ODU); by the Schweizersche Nationalfonds (Basel); by the
Commonwealth of Virginia; and by the Ministry of Science, Culture and
Education of Japan (Tohoku).

\pagebreak

\begin{table}[b] 
\caption{Cross sections for carbon from E143 in nb/GeV/sr (per nucleon)
and $R$ extracted from this experiment and NMC deuterium data  with 
statistical and systematic errors.}
\label{tb:sigma}
\begin{tabular}{ccccc}
   x & $Q^2$ (GeV/c$^2)$ & $\epsilon$ & $\sigma ~\pm$ stat $\pm$sys  & $R ~\pm$ stat $\pm$sys  \\
  0.0325 &  1.32 &   0.474 & 61.91 $\pm$  .46 $\pm$ 3.7 &  0.45 $\pm$   0.01 $\pm$   0.07\\
  0.0375 &  1.47 &   0.519 & 60.81 $\pm$  .50 $\pm$ 3.6 &  0.51 $\pm$   0.02 $\pm$   0.09\\
  0.0450 &  1.67 &   0.578 & 62.82 $\pm$  .53 $\pm$ 2.9 &  0.40 $\pm$   0.01 $\pm$   0.10\\
  0.0550 &  1.90 &   0.641 & 65.64 $\pm$  .62 $\pm$ 2.3 &  0.28 $\pm$   0.01 $\pm$   0.09\\
  0.0650 &  2.11 &   0.692 & 66.76 $\pm$  .66 $\pm$ 2.1 &  0.29 $\pm$   0.02 $\pm$   0.10\\
  0.0750 &  2.29 &   0.734 & 69.61 $\pm$  .73 $\pm$ 2.4 &  0.18 $\pm$   0.02 $\pm$   0.11\\
  0.0850 &  2.46 &   0.767 & 70.23 $\pm$  .73 $\pm$ 1.8 &  0.26 $\pm$   0.03 $\pm$   0.10\\
  0.0950 &  2.60 &   0.795 & 72.03 $\pm$  .70 $\pm$ 2.2 &  0.25 $\pm$   0.03 $\pm$   0.14\\
  0.1050 &  2.73 &   0.818 & 74.46 $\pm$  .71 $\pm$ 2.1 &  0.17 $\pm$   0.03 $\pm$   0.13\\
\end{tabular}		    
\end{table}


\begin{table}[b]
\caption{  Coefficients to 6-parameter fits a, b and c for R1998 
 with the corresponding $\chi^2$/df for 231 degrees of freedom .}
\label{tb:coef}
\begin{tabular}{cccccccc}
 fit & 1 & 2 & 3  & 4 & 5 & 6 & $\chi^2/df$ \\
\hline
a &     0.0485 &     0.5470 &     2.0621 &    -0.3804 &     0.5090 &    -0.0285 &  0.9 \\ 
b &     0.0481 &     0.6114 &    -0.3509 &    -0.4611 &     0.7172 &    -0.0317 &  0.9 \\ 
c &     0.0577 &     0.4644 &     1.8288 &    12.3708 &   -43.1043 &    41.7415 &  1.0 \\ 
\end{tabular}
\end{table}

\pagebreak

\end{document}